\documentclass[12pt,preprint]{aastex}
\usepackage{graphics,graphicx}

\received{2002 November 22}
\begin{document}

\title{Refining Chandra/ACIS Subpixel Event Repositioning Using a Backside
Illuminated CCD Model}

\author{Jingqiang Li\altaffilmark{1}, Joel H. Kastner\altaffilmark{1}, Gregory Y. Prigozhin\altaffilmark{2},
Norbert S. Schulz\altaffilmark{2}}

\altaffiltext{1}{Chester F. Carlson Center for Imaging Science, Rochester Institute of
Technology, 54 Lomb Memorial Dr., Rochester, NY 14623; JL's
email: jxl7626@cis.rit.edu}
\altaffiltext{2}{Center for Space Research, Massachusetts Institute of Technology,
Cambridge, MA 02139}

\begin{abstract}

Subpixel event repositioning (SER) techniques have been
demonstrated to significantly improve the already unprecedented
spatial resolution of Chandra X-ray imaging with the Advanced CCD
Imaging Spectrometer (ACIS). Chandra CCD SER techniques are based
on the premise that the impact position of events can be refined,
based on the distribution of charge among affected CCD pixels.
ACIS SER models proposed thus far are restricted to corner split
(3- and 4-pixel) events, and assume that such events take place at
the split pixel corners. To improve the event counting statistics,
we modified the ACIS SER algorithms to include 2-pixel split
events and single pixel events, using refined estimates for photon
impact locations. Furthermore, simulations that make use of a
high-fidelity backside illuminated (BI) CCD model demonstrate that
mean photon impact positions for split events are energy dependent
leading to further modification of subpixel event locations
according to event type and energy, for BI ACIS devices. Testing
on Chandra CCD X-ray observations of the Orion Nebula Cluster
indicates that these modified SER algorithms further improve the
spatial resolution of Chandra/ACIS, to the extent that the
spreading in the spatial distribution of photons is dominated by
the High Resolution Mirror Assembly, rather than by ACIS
pixelization.
\end{abstract}


\section{Introduction}
Shortly after its launch in 1999, the Chandra X-ray Observatory
(CXO) began delivering X-ray images at exceptional spatial
resolution. Now in routine operation, CXO continues to push the
frontiers of X-ray astronomical imaging. When used in combination
with the Advanced CCD Imaging Spectrometer (ACIS), CXO offers a
potent combination of sub-arcsecond spatial resolution and
moderate ($E/\Delta E \sim 10-50$)  spectral resolution. However,
with its $\sim0''.49$ pixels, ACIS does not fully sample the point
spread function (PSF) of the CXO High Resolution Mirror Assembly
(HRMA). Therefore, in principle, there is further room for
improvement in the resolving power of CXO/ACIS.

Tsunemi et al.\ (2001) have taken advantage of knowledge of X-ray
event charge distributions among CCD pixels and the (subpixel)
telescope pointing history, both of which are included as standard
supporting data for an CXO observation, to generate such an
improvement. Park et al.\ (2002) and Kastner et al.\ (2002) made
use of the basic Tsunemi et al.\ (2001) method to obtain superior
spatial resolution in CXO/ACIS imaging of the supernova remnant SN
1987A and of planetary nebulae, respectively. This work
demonstrates the potential impact of efforts to maximize the
spatial resolving power of CXO/ACIS.

Kastner et al.\ (2002) employed refinements to the Tsunemi et al.\
(2001) method that appear to better capitalize on its potential.
In this paper, we fully explore these refinements, and describe
additional modifications that exploit the energy resolution of
ACIS.

When employed as the focal plane imaging array for CXO, ACIS
collects incident X-ray photons in photon counting mode, which
implicitly assumes that there is at most one photon in a \(
3\times 3 \) subarray in one frame. ACIS registers individual
incoming photons individually in an event list, which records
spatial and spectral information, as well as the event grade. The
grade indicates charge split morphology%
\footnote{An ACIS keyword, FLTGRADE, gives the charge split morphology,
i.e., how many and which neighboring pixels exceed a specified split threshold.
Three groups of split events, i.e. 2-pixel, 3-pixel and 4-pixel split
events, have different average shifts respectively, but we do not distinguish
the difference for a given group in different directions. In other words, events
with FLTGRADE of 11, 22, 104 and 208 all are 4-pixel split events but split
to different corners; the absolute offsets are the same for those events.
} in an isolated \( 3\times 3 \) pixel island centered at the event
pixel.

ACIS processing tools implemented by the Chandra X-ray Center
(CXC) assume that all events have the same photon impact positions
(PIPs), i.e., at the event pixel centers. However, because the
charge cloud size is very small compared with the ACIS CCD pixel
size (Tsunemi et al.\ 1999), the photon impact positions for split
events will be close to the split boundaries instead of the pixel
centers, offering the opportunity of subpixel event repositioning
(SER) derived from the event charge distribution (grades). In
addition, Chandra's slow but intentional dither motion during a
pointed observation moves the target across the detector surface,
in principle allowing full sampling of the PSF of the HRMA, which
otherwise would be subcritically sampled by ACIS.

Tsunemi et al.\ (2001) first proposed an SER algorithm; their
algorithm uses corner split events only. They assume that, for
3-pixel or 4-pixel split events, the actual photon impact
positions are the split corners instead of the pixel centers. So
the algorithm's implementation consists of shifting events by
one-half pixel along both pixel sides towards the split corner
in chip coordinates%
\footnote{There are three fundamental coordinate systems in CXO
event list, i.e., chip coordinates, detector coordinates and sky
coordinates. The conversion among them is unique. Refer to
McDowell J. (2001) for detail. }, then projecting the new location
into the sky coordinates according to the chip orientation and the
spacecraft roll angle. They also predicted that solely based on
corner split events, the knowledge of photon impact positions can
improve by roughly a factor of 10. They conclude that X-ray images
constructed from repositioned corner split events only are almost
free from degradation by the CCD pixel sampling.

However, there is a relatively small percentage of corner split
events in a typical CXO observation. Tsunemi et al.\ (2001) and
Kastner et al.\ (2002) note that corner split events only
constitute about 4\% to 16\% of total events, depending on the
source spectrum and CCD type employed. This is the case even for
BI devices, which should generate more charge-split events than
frontside-illuminated devices. Thus the improvement of spatial
resolution due to the Tsunemi et al.\ (2001) SER algorithm is
{}``at the cost of low efficiency'' and suffers for faint sources.
In addition, since the PSF of the HRMA limits the spatial
resolution, the spatial resolution of Chandra would reach its
maximum, i.e., be telescope limited, so long as we critically
oversample the PSF, e.g., sampling at $0''.25$.

Mori et al.\ (2001) modified the Tsunemi et al.\ (2001) SER method
by adding 2-pixel split events and single pixel events, in order
to improve the statistics. Both the Tsunemi et al.\ and Mori et
al.\ methods assume that all the corner split events take place
precisely at the split pixel corners, while 2-pixel split events
occur exactly at the centers of split boundaries. Physical CCD
models (Prigozhin et al.\ 2002) have demonstrated, however, that
corner split events can be formed even for photon impact somewhat
far from the corners, where the distance is a function of photon
energy. These simulations indicate that the assumed positions of
split events can be refined via a physical model of the CCD-photon
interaction.

\section{Modifying Subpixel Event Repositioning (SER) by Expanding Event Selection
Criteria}

To overcome small number statistics problems and thereby improve
on the SER proposed by Tsunemi et al.\ (2001), we want to use all
13 {}``viable'' event grades%
\footnote{The grade here means FLTGRADE defined by ACIS
instrument. Although there are 256 different grades, events with
charge distributed over more than 4 pixels are most probably
formed by noise, like cosmic rays. The 13 grades corresponding to
events covering 4 pixels or less can be divided into three
sub-groups, i.e., single pixel event, 2-pixel split event, and
corner split event. Those 13 grades account for approximately 95\%
of total events for typical X-ray sources. Refer to Kastner et
al.\ (2002) for the 13 selected grades.}. Upon analysis of the
ACIS simulations (see sec. \ref{sec:sec3}), we found that most
single pixel events occur near the pixel centers, and constrained
in an area slightly smaller than an ACIS pixel. Two-pixel split
events are generated by photons that are absorbed near the centers
of split boundaries, while the impact positions of corner split
events are limited to the smallest area, close to pixel corners.
Because the charge cloud size is very small compared with ACIS
pixel size, single-pixel events will have the biggest position
uncertainty in both directions along pixel sides, and corner split
events have the smallest uncertainty among all the events in both
directions, while 2-pixel split events have relatively small
uncertainty in the direction perpendicular to split boundary, and
have uncertainties similar to those of single-pixel events in the
direction parallel to split boundary.

\begin{figure}
{\centering \resizebox*{4in}{!}{\includegraphics{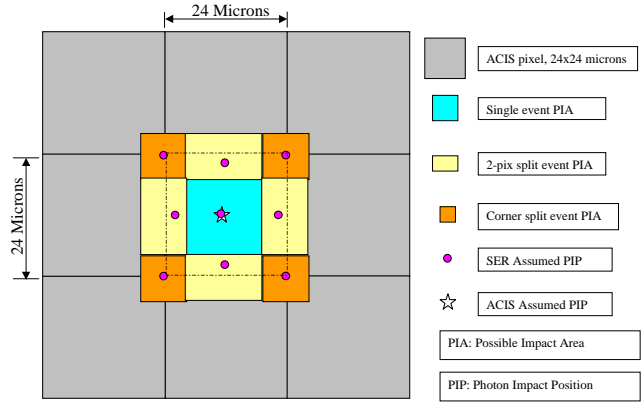}} \par}

\caption{Schematic illustration of the subpixel event
repositioning algorithm. A 3 $\times$ 3 pixel island is displayed,
where the central pixel is the event pixel. The solid circles
represent SER assumed PIPs according to the event grades, while
the shaded areas are the possible photon impact areas, for
different event sub-groups.} \label{subpixel}
\end{figure}

Therefore, in our implementation of SER, we have modified the
Tsunemi et al.\ (2001) model by adding 2-pixel split events and
single pixel events. We assume that corner split events take place
at the split corners instead of event pixel centers (as also
assumed by Tsunemi et al.\ 2001), and 2-pixel split events occur
at the centers of split boundaries, 0.366 pixel away from the
pixel centers. Single pixel event PIPs remain at the event pixel
centers, as we have no way to improve these PIPs based on charge
distribution. The 0.366 pixel offset for 2-pixel split events was
determined empirically by minimizing the PSFs of point source
images in on-orbit ACIS BI CCD data (see sec. \ref{sec:sec4}). We
refer to this modified algorithm as {}``static''
(energy-independent) SER. The algorithm's schematics can be found
in figure \ref{subpixel}, in which the pixel island and assumed
photon impact positions are displayed.

\clearpage
\clearpage

\begin{figure}
{\centering \resizebox*{5in}{!}{\includegraphics{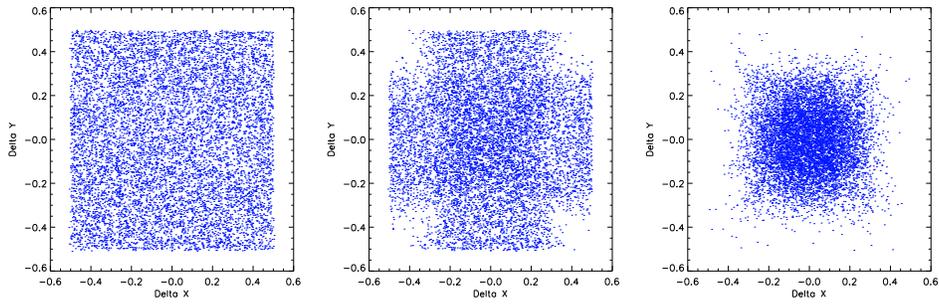}} \par}

\caption{The difference between actual photon impact position
(PIP) and processed event assumed location for 1.74 keV events, in Chip
coordinates. Left panel: ACIS assumed PIP; middle panel:
correction using corner events only; right panel: modified SER
correction. The panels are in units of pixels.} \label{half_shift_compare}
\end{figure}

Figure \ref{half_shift_compare} shows the improvement enabled by
the modified SER, using the back-illuminated CCD simulated data at
the energy of 1740 eV. In this three-panel plot, we show the
differences between actual PIPs and various models for PIPs in
chip coordinates, for all three subgroup events. The plot axes are
in pixel units, i.e., 0.5 difference indicates photons interacted
near the pixel boundaries. The left panel shows the difference of
actual PIP with PIPs assumed to lie at event pixel centers; one
can see the expected uniform random distribution within the pixel.
The middle panel is the difference after corner-event-only SER
correction (i.e., the Tsunemi et al.\ 2001 model). A big
improvement for events that occur near corners can be seen.
However, due to the small proportion of corner split events, there
is no correction for most events, even those occurring near
boundaries. The right panel shows the difference after the static
SER correction. The PIP difference for the static SER model is
more compact than for the Tsunemi et al.\ (2001) SER model,
indicating that the modified PIPs are closer to their real
locations. The potential for image quality improvement, using
static SER, is apparent.

\clearpage
\clearpage

\section{Further Modifications to SER Based on ACIS BI CCD Simulations}
\label{sec:sec3}

\begin{figure}
{\centering \resizebox*{3in}{!}{\includegraphics{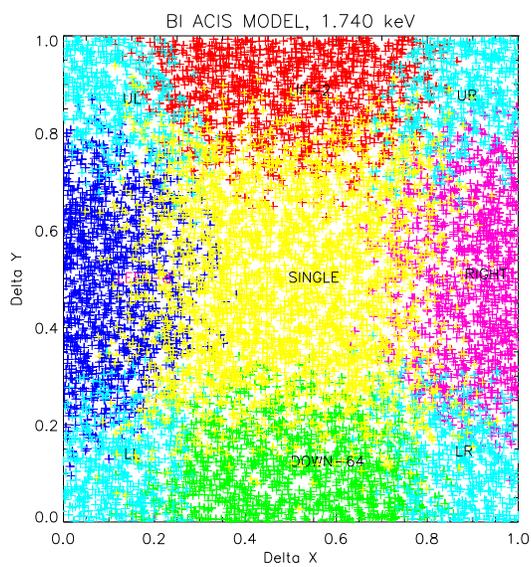}} \par}

\caption{The photon impact positions for 3 subgroups of 13
{}``viable'' event grades. Plus signs stand for the PIPs of
2-pixel events within a pixel, while triangles represent the PIPs
of corner (3- or 4- pixel) split events. The crosses are the PIPs
of single pixel events. All the photons have energy of 1.74 keV.}
\label{splitevent}
\end{figure}

Simulations show that not all corner split events take place at
the pixel corners; nor do the 2-pixel split events occur at the
centers of split boundaries (figure \ref{splitevent}). Instead,
the corner split events can take place as far as 0.3 pixel away
from the split corners. In addition, within a certain small area
of a pixel, all three kinds of events could be generated;
furthermore, the size of the area that produces specific event
charge splits changes according to photon energy. Therefore the
critical question for further SER modification is how best to
determine the shifts for the split events. For this reason, we
have used a physical model of the BI CCD (Prigozhin et al.\ 2002)
to determine the appropriate position shifts, according to photon
energies.
\clearpage
\clearpage

The simulator we used is a Monte Carlo model of a
backside-illuminated ACIS CCD (CCID-17). The simulator can only be
used for monochromatic simulations; the code itself randomly
generates photon impact positions with subpixel accuracy and
simulates where the electron clouds were formed and how the charge
was spread across the pixels. The output of the simulation
includes subpixel photon impact locations and the signal
amplitudes in the pixels of the 3 $\times$ 3 event island. Thus
the simulation with the BI CCD model enables us to analyze how
charge is split when photon impact positions are close to
boundaries.

We performed simulations consisting of 10,000 photons at each
energy from 300 eV to 12 keV, with an energy step of 100 eV.
Because of the attenuation-length jump at the silicon edge near
1800 eV, we lowered the energy step to 10 eV from 1800 eV to 1900
eV. For each simulation, we calculated the event grade percentages
for single pixel events, 2-pixel split events, and corner split
events; and the average subpixel position shifts for 2-pixel split
events, 3-pixel split events, and 4-pixel split events.

\clearpage
\begin{figure}
{\centering \resizebox*{4in}{4in}{\includegraphics{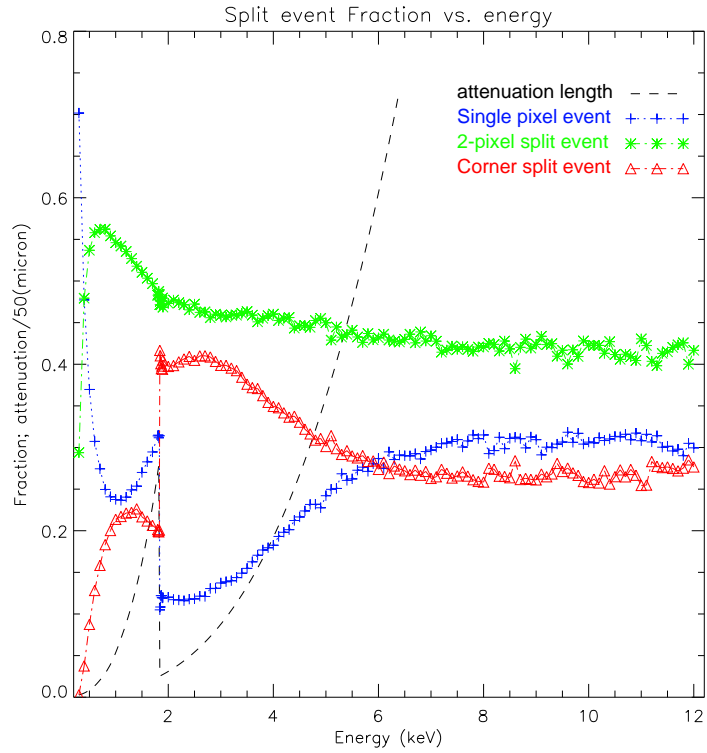}} \par}

\caption{The fraction of different event grades versus photon
energy, with the X-ray attenuation length in silicon overplotted.
Note that the attenuation length is in units of 50 microns. } \label{split_percentage}
\end{figure}
\clearpage

\begin{figure}
{\centering \resizebox*{4in}{4in}{\includegraphics{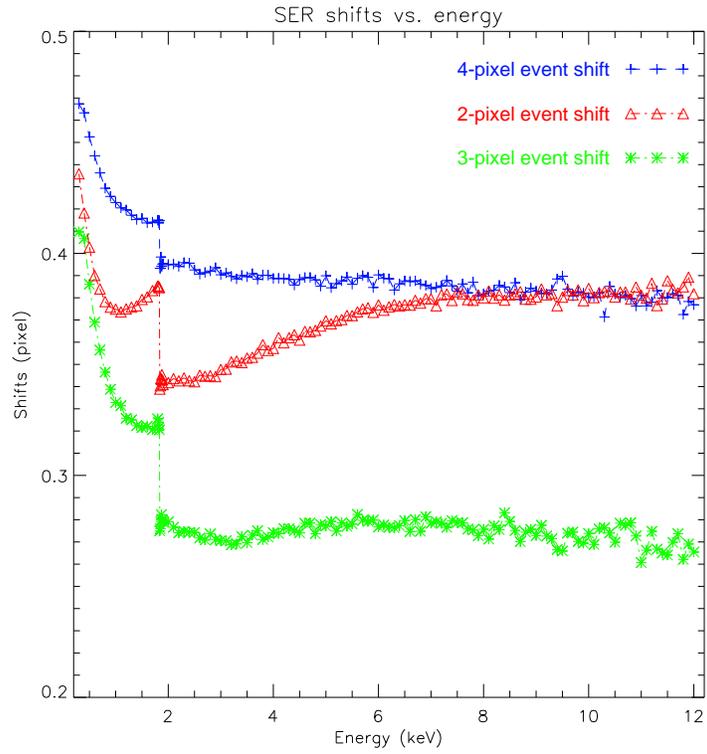}} \par}

\caption{The mean shifts from pixel centers of the 3 groups of
split events, according to the photon energy.}
\label{shift_energy}
\end{figure}

Figure \ref{split_percentage} shows the event percentage as a
function of photon energy, while figure \ref{shift_energy} shows
the mean shift in position for different split event types. Note
that at low energy ($<$ 2 keV), both subgroup event percentage and
SER shifts depend sensitively on energy. The two figures clearly
show the jumps at the silicon absorption edge. Above 6 keV, the 3
subgroup event percentages and PIP shifts are insensitive to
energy. This can be explained by the fact that, for photons with
energy exceeding 6 keV, the characteristic penetration depth
becomes comparable to or larger than the thickness of the ACIS BI
CCD, which is only 45 microns.

\clearpage
\clearpage

The improvement in PIP determination benefits from applying the
average energy-dependent shifts for different split event groups.
Based on the simulations, we calculated these average shifts for
different kinds of split events. Then, we add an offset to the
photon impact location in chip coordinates according to charge
split morphology and photon energy. We refer to this SER
modification as {}``energy-dependent'' SER.

\begin{figure}
{\centering \resizebox*{5in}{!}{\includegraphics{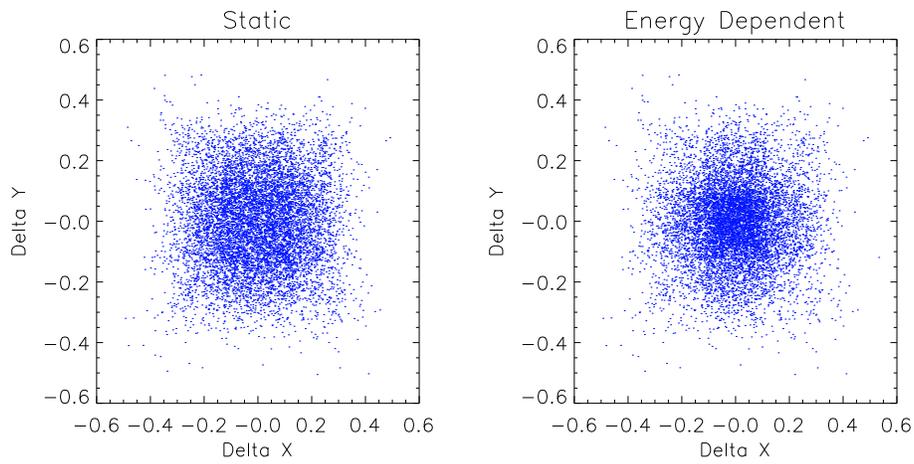}} \par}

\caption{The differences between SER assumed photon impact
positions (PIPs) and actual PIPs for {}``static'' (left) and
energy-dependent (right) SER models. All the simulated photons are
at the energy of 1.74 keV. The cross like structure in
the right panel comes from split events, for which we apply the
same mean offsets for all events with the same FLTGRADE, while the
{}``halo'' is generated by single-pixel events, for which impact
positions are more uncertain.} \label{compare}
\end{figure}

Figure \ref{compare} shows results for the static and
energy-dependent SER algorithms, at an energy of 1740 eV. For
comparison, we include the right-most panel in figure
\ref{half_shift_compare} as the left panel. The plot shows the
differences between actual and calculated photon impact positions
after applying SER corrections. The right panel shows the PIP
differences for energy-dependent SER, i.e., by relocating events
according to both event grade and energy, with shifts calculated
from a look up table derived from the results illustrated in
figure \ref{shift_energy}. Clearly, one can see the improvement using
energy-dependent SER.

\clearpage
\clearpage

\section{Application of Energy-Dependent SER}
\label{sec:sec4}

We have applied both modified SER algorithms to real Chandra/ACIS
data, as an evaluation tool. The tested data is an observation of
Orion Nebula Cluster (ONC) which was obtained in 1999 using the
back illuminated CCD ACIS-S3 \emph{}(Schulz et al.\ 2001). The
algorithm implementation is as follows:

\begin{enumerate}
\item Remove event position randomization that was applied by CXC
standard ACIS processing.
\item Filter out the 13 {}``viable''
event grades (including single-pixel events) according to the
FLTGRADE.
\item From the look up table derived from results in
figure \ref{shift_energy}, calculate the offset according to
photon grade and energy, and add the offset in the appropriate
direction(s), in chip coordinates, for every event.
\item Project the new chip coordinates for relocated events%
\footnote{Including single-pixel events whose PIPs were not changed
by SER algorithms.} to sky coordinates according to the roll angle
of the spacecraft and the orientation of the employed CCD.
\item
Reconstruct each point source from the old and new sky
coordinates, fit the source with a two dimensional Gaussian
function, and calculate the full width half
maxima (FWHMs) before and after applying SER.
\end{enumerate}

\subsection{Results}

\begin{figure}
{\centering \resizebox*{4in}{!}{\includegraphics{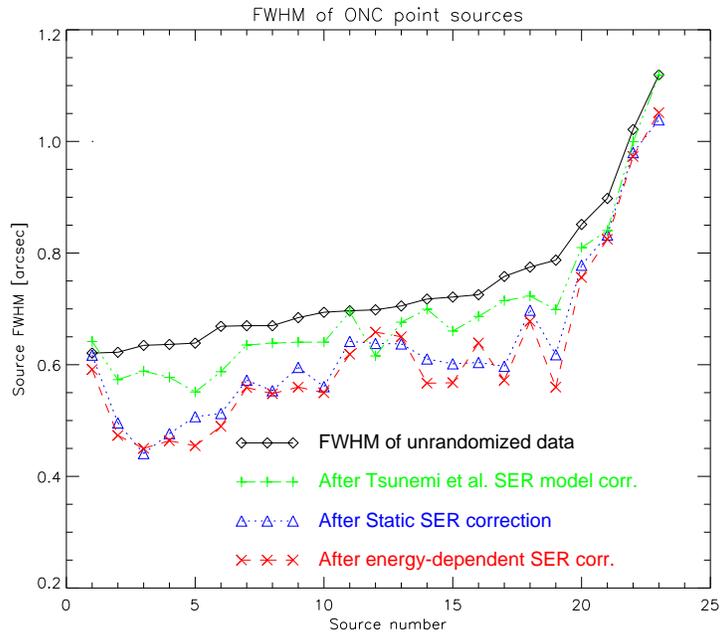}} \par}
\caption{FWHM of BI ONC point like sources before
and after applying various SER algorithms described in this paper. }
\label{fwhm_imp}
\end{figure}

Based on the above steps, we have plotted the FWHM of 22 bright 
point-like sources in BI ONC data. The
sources are chosen to represent range in count rate from $0.0052$
to $0.2791$ s$^{-1}$, and in off-axis angle from $2''.72$ to
$271''.6$. Figure \ref{fwhm_imp} and table \ref{tbl1} show that
after applying SER algorithm to these data, all SER
algorithms (sec. \ref{sec:sec3}) improved the FWHM for every
source (except that source 1 has no improvement after applying the
Tsunemi et al.\ [2001] method). The abscissa axis is source
number, sorted with the FWHM of original point source, before
applying SER but after removing randomization. Furthermore, 17 out
of 22 sources have better (smaller) FWHM for energy-dependent SER
than static SER, showing that energy-dependent SER has better
capability to improve Chandra/ACIS PSF function. For comparison,
we include the FWHMs after applying Tsunemi et al.\ (2001) SER
model. As expected, both modified SER algorithms demonstrate
better improvement than this original model.

\clearpage
\clearpage

\begin{figure}
{\centering
\resizebox*{4in}{!}{\includegraphics{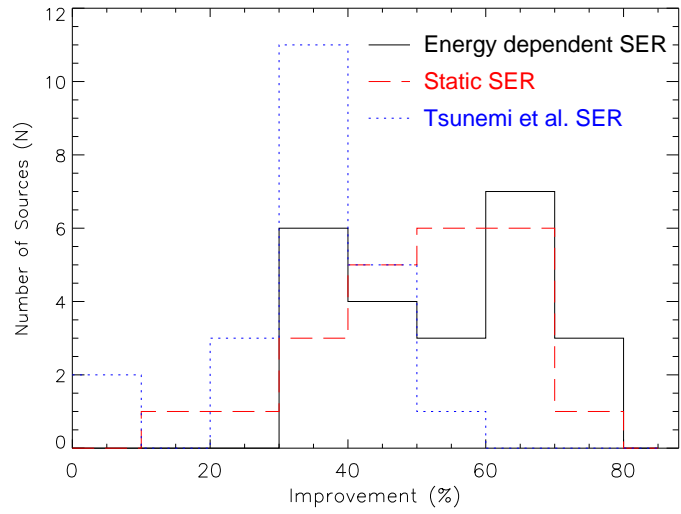}} \par}

\caption{Comparison of image FWHM improvements using Tsunemi et
al.\ model, static and energy-dependent SERs on BI CXC ONC data.}
\label{improve}
\end{figure}

In order to evaluate the achievement of the SER algorithm, we
follow Tsunemi et al. (2001) and use the FWHM of a point source to
define the improvement. Assuming that {$\mit F_{B}$} and {$\mit
F_{A}$} are the FWHMs of a source before and after applying SER,
respectively, the improvement $\Delta$ is defined as:
\begin{displaymath}
\Delta\,=\,\sqrt{{\mit F_{B}^{2}\,-\,F^{2}_{A}}}/F_{B}
\end{displaymath}
Figure \ref{improve} shows this metric of the improvement of the
static and energy-dependent SER techniques. We include Tsunemi et
al.\ (2001) model for comparison. As expected, the plot shows that
most sources have superior FWHM improvement for energy-dependent
SER than for static SER; only 5 of 22 display inferior improvement
(figure \ref{fwhm_imp}). Improvements in FWHM range from 40\% to
70\%, for most sources, with the improvement somewhat dependent on
off-axis angle, for the ONC source sample included here. There are
two possible reasons why the improvements in FWHM for the
energy-dependent SER, relative to static SER, are somewhat smaller
than might be anticipated from the improvement in PIPs (figure
\ref{shift_energy}): \begin{enumerate} \item Under either modified
SER method, the HRMA PSF dominates the spatial distributions of
source photons. That is, CXO/ACIS images are likely
telescope-limited utilizing either static or energy-dependent SER,
especially at large ($\geq$ 2 arcminutes) off-axis angle. \item
There remain uncertainties in the CCD model, in particular, the
grade branching ratios predicated by the model prevent further
improvement in source FWHM, using energy-dependent SER.
\end{enumerate}
We are conducting further experiments with SER-processed ACIS data
obtained for the ONC, to attempt to distinguish between these
possibilities.

\clearpage
\clearpage

\section{Summary}
We have conducted a study of potential improvements to subpixel
event repositioning (SER) for CXO/ACIS data. We
formulate modified SER algorithms at two levels of improvement:
(1) inclusion of single-pixel events and two-pixel split events
({}``static'' SER); (2) accounting for the mean energy dependence
of differences between apparent and actual photon impact
positions, based on the results of BI CCD simulations
({}``energy-dependent'' SER).

We find that both static and energy-dependent SER produce
improvements in spatial resolution over those possible using an
earlier static SER algorithm employing only corner-split events
(Tsunemi et al.\ 2001). The potential improvement in image FWHM is
$\sim$50\% using either of the {}``modified SER'' algorithms
described here, with energy-dependent SER producing a marginally
superior result. The relatively small improvement observed for
energy-dependent SER, relative to static SER, suggests that under
either method, the HRMA --- rather than ACIS pixelization ---
dominates image FWHM.

SER techniques only take into account the properties of photon
charge-splitting within CCD pixels, and do not depend on the
characteristics of the telescope (in particular, its PSF).
Therefore, SER is applicable to both compact and extended sources
(Kastner et al.\ 2002, Li et al.\ 2003). However, deconvolution
methods have been developed in recent decades for optical and IR
astronomical imaging, to correct for the blurring due to telescope
PSF. Burrows et al.\ (2000) used one technique, Maximum
Likelihood, to deconvolve  the ACIS-S image of SN 1987A. At
present, multiscale deconvolution methods are being explored,
which are more suitable to process Poisson-distributed data, and
therefore may be better applicable to X-ray imaging (Willett et
al.\ 2003, Esch et al. 2003). By combining SER techniques and such
multiscale deconvolution methods, one can expect the best possible
spatial resolution from Chandra/ACIS imaging.

\acknowledgements{We acknowledge helpful discussions with Leisa
Townsley, Patrick Broos, and Herman Marshall, and helpful
suggestions from the referee, Scott Wolk. This research was
supported by NASA/CXO grant G02-3009X to RIT.}

\begin{table}
\caption{Information for the sources appearing in figure
\ref{fwhm_imp}.}\label{tbl1}
\begin{tabular}{cccccccccc}
\hline
\footnotesize{}
  & & & & $\theta^{d}$ & $R^{e}$ & $FHWM_{o}^{f}$ &\multicolumn{3}{c}{Improvement (\%)} \\
  Source & $\alpha (2000)^{a}$ & $\delta (2000)^{b}$ & SSN$^{c}$ & (arcsec) &  (counts/s) & (arcsec) & Tsunemi & Static & E-Dep.$^{g}$ \\
\hline
\hline
       1 &   18.36 & 22 37.38 & 91 &   55.20 &      0.014  &  0.62&        ...&      10.73 &      30.39 \\
       2 &   15.63 & 22 56.44 & 55 &   28.48 &      0.023  &  0.62&      38.86&      60.45 &      64.92 \\
       3 &   17.00 & 22 32.95 & 76 &   51.15 &      0.005  &  0.63&      37.46&      71.93 &      70.50 \\
       4 &   15.26 & 22 56.83 & 48 &   30.30 &      0.012  &  0.64&      42.01&      66.26 &      68.48 \\
       5 &   16.46 & 23 22.89 & C  &    2.72 &      0.279  &  0.64&      50.53&      60.82 &      70.22 \\
       6 &   15.77 & 23 09.86 & E  &   15.10 &      0.077  &  0.67&      47.78&      64.23 &      68.14 \\
       7 &   14.32 & 23 08.31 & 24 &   32.67 &      0.008  &  0.67&      31.72&      51.99 &      55.07 \\
       8 &   12.29 & 23 48.06 &  7 &   64.74 &      0.009  &  0.67&      30.25&      56.36 &      57.53 \\
       9 &   17.94 & 22 45.42 & 85 &   45.10 &      0.092  &  0.68&      35.27&      49.35 &      57.48 \\
      10 &   15.82 & 23 14.19 &  A &   11.27 &      0.040  &  0.69&      38.52&      58.92 &      61.03 \\
      11 &   21.03 & 23 48.00 & 108 &  76.27 &      0.035  &  0.70&       7.34&      38.91 &      45.95 \\
      12 &   14.55 & 23 16.01 & 30 &   26.80 &      0.009  &  0.70&      47.19&      40.62 &      33.31 \\
      13 &   17.06 & 23 34.09 & 78 &   16.37 &      0.012  &  0.71&      28.46&      42.86 &      38.78 \\
      14 &   15.34 & 22 15.47 & 50 &   69.09 &      0.012  &  0.72&      22.37&      52.59 &      61.31 \\
      15 &   19.20 & 22 50.63 & 97 &   54.57 &      0.024  &  0.72&      40.20&      55.14 &      61.69 \\
      16 &     ...  &     ... & ... &  97.24 &      0.009  &  0.73&      32.36&      55.39 &      47.33 \\
      17 &   14.91 & 22 39.14 & 40 &   48.54 &      0.007  &  0.76&      33.43&      61.63 &      65.61 \\
      18 &     ...  &     ... & ... &  85.62 &      0.018  &  0.77&      35.98&      43.51 &      48.45 \\
      19 &   15.97 & 23 49.70 & 60 &   27.22 &      0.014  &  0.79&      46.07&      61.97 &      70.31 \\
      20 &     ...  &     ... & ... &  115.42 &      0.012 &  0.90 &      35.24&      37.51 &      39.57 \\
      21 &     ...  &     ... & ... &  136.75 &      0.023 &  1.02 &      20.59&      28.17 &      30.39 \\
      22 &     ...  &     ... & ... &  120.50 &      0.009 &  1.12 &       3.66&      37.27 &      34.26 \\
\hline
\end{tabular}
Notes.---
a. Right ascension for all sources is at $5^{h}35^{m}$; values in table are in units of
seconds. \\
b. Declination for all sources is at $-5^{\circ}$; values in table are in units of
arcminutes and arcseconds. \\
c. SSN stands for Schulz et al.\ (2001) source number.\\
d. $\theta$ = off-axis angle.\\
e. $R$ = source count rate.\\
f. $FWHM_{o}$ is FHWM after removing randomization but before applying SER.\\
g. E-dep. represents result from energy-dependent SER.\\

\end{table}

\end{document}